\documentclass[12pt]{article}
\usepackage{a4wide}
\usepackage{latexsym}
\usepackage{axodraw}
\usepackage{amsmath}
\usepackage{amsfonts}
\usepackage{amscd}
\usepackage{cite}

\usepackage{pslatex}
\usepackage{graphicx}
\usepackage[latin1]{inputenc}
\usepackage[T1]{fontenc}

\newcommand{\bq}{\begin{eqnarray}}
\newcommand{\eq}{\end{eqnarray}}
\newcommand{\eps}{\varepsilon}

\begin{document}

\thispagestyle{empty}

\begin{flushright}
  MITP/13-016
\end{flushright}

\vspace{1.5cm}

\begin{center}
  {\Large\bf The two-loop sunrise graph with arbitrary masses\\
  }
  \vspace{1cm}
  {\large Luise Adams ${}^{a}$, Christian Bogner ${}^{b}$ and Stefan Weinzierl ${}^{a}$\\
  \vspace{1cm}
      {\small ${}^{a}$ \em PRISMA Cluster of Excellence, Institut f{\"u}r Physik, }\\
      {\small \em Johannes Gutenberg-Universit{\"a}t Mainz,}\\
      {\small \em D - 55099 Mainz, Germany}\\
  \vspace{2mm}
      {\small ${}^{b}$ \em Institut f{\"u}r Physik, Humboldt-Universit{\"a}t zu Berlin,}\\
      {\small \em D - 10099 Berlin, Germany}\\
  } 
\end{center}

\vspace{2cm}

\begin{abstract}\noindent
  {
We discuss the analytical solution of the two-loop sunrise graph with arbitrary non-zero masses in two space-time dimensions.
The analytical result is obtained by solving a second-order differential equation.
The solution involves elliptic integrals and in particular the solutions of the corresponding homogeneous differential equation are given
by periods of an elliptic curve.
   }
\end{abstract}

\vspace*{\fill}

\newpage

\section{Introduction}
\label{sec:intro}

The two-loop sunrise graph with non-zero masses is the simplest Feynman integral, which cannot be expressed in terms of
multiple polylogarithms.
It has already received considerable attention in the 
literature 
\cite{Broadhurst:1993mw,Berends:1993ee,Bauberger:1994nk,Bauberger:1994by,Caffo:1998du,Laporta:2004rb,Groote:2005ay,Groote:2012pa,Bailey:2008ib,MullerStach:2011ru}.
The analytical solution for the equal mass case has been discussed in \cite{Laporta:2004rb}.
Up to now, less is known for the unequal mass case.
The state of the art in the unequal mass case can be summarised as follows:
First of all, expansions around special points, like zero momentum squared, threshold or pseudo-thresholds
are known 
\cite{Berends:1997vk,Davydychev:1999ic,Caffo:1999nk,Caffo:2001de,Groote:2000kz,Onishchenko:2002ri}.
Furthermore, representations of the full integral in terms of Lauricella functions
\cite{Berends:1993ee,Bauberger:1994nk,Bauberger:1994by,Bauberger:1994hx}
or one-dimensional integral representations involving Bessel functions
\cite{Groote:2005ay,Groote:2012pa}
are also known.
For practical purposes, numerical evaluations are available
\cite{Caffo:2002ch,Pozzorini:2005ff,Caffo:2008aw}.

The two-loop sunrise integral with non-zero masses is relevant for precision calculations in electro-weak physics \cite{Bauberger:1994nk}, 
where non-zero masses naturally occur.
In addition, the two-loop sunrise integral with non-zero masses appears as a sub-topology in many advanced higher-order calculations, like
the two-loop corrections to top-pair production or in the computation of higher-point functions in massless theories \cite{Paulos:2012nu,CaronHuot:2012ab}.
It is therefore desirable to understand this integral in detail, as such an understanding will pave the way for more complicated processes.
As already mentioned, the two-loop sunrise integral provides the simplest example of an integral which cannot be expressed in terms of multiple polylogarithms.
Therefore this integral is the ideal playground to learn more about transcendental functions occuring in Feynman integrals beyond multiple polylogarithms.

In this paper we discuss the two-loop sunrise integral with unequal non-zero masses in two space-time dimensions.
The restriction to two space-time dimensions allows us to focus on the essential difficulties and avoids
some entanglements with complications which we know how to handle.
In two space-time dimensions the integral is finite and depends -- as we will see later -- only on the second graph polynomial, but not 
on the first graph polynomial.
Working in two space-time dimensions removes therefore the issues of ultraviolet divergences and the dependence on the first graph polynomial.
Of course we are interested in the end in the result around four space-time dimensions.
Using dimensional recurrence relations \cite{Tarasov:1996br,Tarasov:1997kx}, 
the result in two space-time dimensions can be related to the result in four space-time dimensions.

Recently, it emerged that methods from algebraic geometry can be very useful for the computation of Feynman integrals.
For example, it has been discovered that the two-loop sunrise integral with unequal masses in two space-time dimensions 
satisfies an ordinary second-order linear differential equation \cite{MullerStach:2011ru}. 
In the equal mass case, such a second-order linear differential equation has been known already for long time \cite{Laporta:2004rb}.
The surprise was that the generalisation to unequal masses does not increase the order of the differential equation.
This can be understood in terms of variations of Hodge structures: The order of the differential equation is related to the dimension
of a certain cohomology group, both numbers being integers.
It turns out that the variation from the equal mass case to the unequal mass case is smooth and in particular the order of the differential equation is a smooth function.
It is clear that a smooth function which takes integer values has to be constant.
In general, Feynman integrals satisfy differential equations of Picard-Fuchs type (ordinary linear differential equations with at most regular singularities) \cite{MullerStach:2012mp}.
In this paper we solve the differential equation for the unequal mass case, similar to what has been done in the simpler case of equal masses in \cite{Laporta:2004rb}.
As a side remark, we would like to mention that inhomogeneous Picard-Fuchs equations have also been discussed recently in the context of mirror symmetry \cite{Laporte:2012hv}.

This paper is organised as follows:
In the next section we discuss the relation between (factorised) differential equations and iterated integrals.
The case where the differential operator factorises into linear factors will lead to multiple polylogarithms.
In this paper we go beyond linear factors and we are interested in the case, where the differential equation contains an irreducible second-order differential operator.
To give a simple and concrete example for the approach by differential equations 
we consider in section~(\ref{sec:one_loop}) as a warm-up exercise the differential equation for the one-loop two-point function and its solution.
We then turn to the two-loop sunrise integral and briefly recall the second-order differential equation for this integral in section~(\ref{sec:definition}).
Section~(\ref{sec:solution}) is the main part of this paper. We discuss the solution of the differential equation by first determining the solutions of the homogeneous equation,
fixing the boundary values and by considering one special solution of the inhomogeneous equation.
Finally section~(\ref{sec:conclusions}) contains our conclusions.
In an appendix we give useful information on how to transform an elliptic integral to the Legendre normal form.
We also give a representation of the solution of the two-loop sunrise integral in terms of multiple sums.

\section{Motivation}
\label{sec:motivation}

We denote by $I_G$ a scalar Feynman integral corresponding to a Feynman graph $G$.
A scalar Feynman integral depends on the external momenta $p_j$ only through the Lorentz invariant quantities
\bq
\label{lorentz_invariant}
 s_{jk} & = & \left( p_j + p_k \right)^2.
\eq
The Feynman integral depends also on the internal masses denoted by $m_i$.
We call the set
\bq
\label{def_set_invariants}
 S & = & \left\{ s_{jk}, m_i^2 \right\}
\eq
the set of kinematical invariants.
Let us pick one variable $t\in S$. We will consider the Feynman integral as a function of $t$ and write $I_G(t)$, the dependence on all other parameters
is implicitly understood.
Recently a systematic method has been developed to find for the Feynman integral a Picard-Fuchs equation \cite{MullerStach:2012mp}.
The Picard-Fuchs equation is an ordinary linear differential equation of a certain order $r$:
\bq
\label{picard_fuchs}
  \sum\limits_{j=0}^r p_j(t) \frac{d^j}{dt^j} I_G(t) & = & \sum\limits_i q_i(t) I_{G_i}(t),
\eq
where $I_{G_i}$ denotes simpler Feynman integrals corresponding to graphs, where one propagator of the original graph has been contracted.
The coefficients $p_j(t)$ of the differential operator and the coefficients $q_i(t)$ appearing on the right hand side are polynomials in $t$.
(Note that in eq.~(\ref{picard_fuchs}) $p_j(t)$ denotes a polynomial in $t$, while in eq.~(\ref{lorentz_invariant}) $p_j$ denotes some external
momentum.)

In the simplest case the differential operator factorises into linear factors:
\bq
\label{linear_factors}
 \sum\limits_{j=0}^r p_j(t) \frac{d^j}{dt^j}
 & = & 
 \left( a_r(t) \frac{d}{dt} + b_r(t) \right)
 ...
 \left( a_2(t) \frac{d}{dt} + b_2(t) \right)
 \left( a_1(t) \frac{d}{dt} + b_1(t) \right).
\eq
Let us denote the homogeneous solution of the $j$-th factor by
\bq
\label{solution_homogeneous_factor}
 \psi_j(t) & = & \exp\left(- \int\limits_0^t ds \frac{b_j(s)}{a_j(s)} \right).
\eq
$b_j(s)/a_j(s)$ is a rational function in $s$ and the integral in eq.~(\ref{solution_homogeneous_factor}) can be expressed in terms
of rational functions and logarithms.
In the case where the Picard-Fuchs operator factorises into linear factors as in eq.~(\ref{linear_factors}), the solution 
of the differential equations is given by iterated integrals
\bq
\label{iterated_solution}
 I_G(t)  & = &
 C_1 \psi_1(t) + \psi_1(t) \int\limits_0^t \frac{dt_1}{a_1(t_1) \psi_1(t_1)} 
                  \left( C_2 \psi_2(t_1) + \psi_2(t_1) \int\limits_0^{t_1} \frac{dt_2}{a_2(t_2) \psi_2(t_2)}
                              ...
 \right. 
 \nonumber \\
 & & \left. 
 ...
                                  \left( C_r \psi_r(t_{r-1}) + \psi_r(t_{r-1}) \int\limits_0^{t_{r-1}} \frac{dt_r}{a_r(t_r) \psi_r(t_r)} 
                                                         \sum\limits_i q_i(t_r) I_{G_i}(t_r)
                              \right) 
           \right).
\eq
The $r$ integration constants are denoted by $C_1$, ..., $C_r$.

In this paper we are interested in cases, where the Picard-Fuchs operators do not factorise into linear factors.
The simplest cases beyond linear factors have an irreducible second-order differential operator.
An example for an irreducible second-order differential operator is given by
\bq
\label{example_diff_eq_K}
 D & = & t \left(1-t^2\right) \frac{d^2}{dt^2} + \left( 1 - 3 t^2 \right) \frac{d}{dt} - t.
\eq
Let us now consider, what happens if we replace in eq.~(\ref{linear_factors}) the $j$-th factor 
\bq
 a_j(t) \frac{d}{dt} + b_j(t)
\eq
by the second-order differential operator
\bq
 a_j(t) \frac{d^2}{dt^2} + b_j(t) \frac{d}{dt} + c_j(t).
\eq
First of all, a second-order differential operator has two independent homogeneous solutions, which we denote
by $\psi_j^{(1)}(t)$ and $\psi_j^{(2)}(t)$.
Once these are known, we can replace in eq.~(\ref{iterated_solution}) the combination
\bq
 C_j \psi_j(t_{j-1}) + \psi_j(t_{j-1}) \int\limits_0^{t_{j-1}} \frac{dt_j}{a_j(t_j) \psi_j(t_j)}
\eq
by
\bq
 C^{(1)}_j \psi^{(1)}_j(t_{j-1}) + C^{(2)}_j \psi^{(2)}_j(t_{j-1}) 
 + \int\limits_0^{t_{j-1}} \frac{dt_j}{a_j(t_j) W_j(t_j)}  
    \left( \psi^{(2)}_j(t_{j-1}) \psi^{(1)}_j(t_j) - \psi^{(1)}_j(t_{j-1}) \psi^{(2)}_j(t_j) \right),
 \nonumber 
\eq
where $W_j(t)$ denotes the Wronski determinant.
The Wronski determinant is defined by
\bq
 W_j(t) & = & \psi_j^{(1)}(t) \frac{d}{dt} \psi_j^{(2)}(t) - \psi_j^{(2)}(t) \frac{d}{dt} \psi_j^{(1)}(t).
\eq
$C^{(1)}_j$ and $C^{(2)}_j$ are two integration constants.
Apart from this substitution the structure of the solution in eq.~(\ref{iterated_solution}) 
as an iterated integral remains unchanged.

The most important ingredients are certainly the two solutions $\psi_j^{(1)}(t)$ and $\psi_j^{(2)}(t)$
of the homogeneous second-order differential equation.
For a second-order differential equation this is a non-trivial task, even in the case where $a_j(t)$, $b_j(t)$ and $c_j(t)$ are
polynomials.
This is in contrast to the linear case, where the generic form in eq.~(\ref{solution_homogeneous_factor})
of the homogeneous solution is known and an algorithm (partial fraction decomposition) is known to perform
the integration in eq.~(\ref{solution_homogeneous_factor}).
Let us come back to the example in eq.~(\ref{example_diff_eq_K}).
This differential operator has as homogeneous solutions the functions
\bq
 \psi^{(1)}(t) = K\left(t\right),
 & &
 \psi^{(2)}(t) = K\left(\sqrt{1-t^2}\right),
\eq
where
$K(x)$ denotes the complete elliptic integral of the first kind
\bq
 K(x)
 & = &
 \int\limits_0^1 \frac{dt}{\sqrt{\left(1-t^2\right)\left(1-x^2t^2\right)}}.
\eq
We will see in the sequel that the differential equation for the two-loop sunrise integral has homogeneous solutions,
which are given by complete elliptic integrals of the first kind times algebraic prefactors.

\section{A warm-up exercise}
\label{sec:one_loop}

We would like to make the formal considerations of the previous section a little bit more concrete.
This is best illustrated by a simple Feynman integral having a first-order differential equation.
Therefore we consider as a warm-up exercise the one-loop two-point functions in two space-time dimensions.
The integral is defined by
\bq
\label{def_one_loop}
 B\left( t \right) & = & 
 \mu^2
 \int\limits_{x_i\ge0} d^2x \; \delta\left(1-x_1-x_2\right)
 \frac{1}{(-t) x_1 x_2 + x_1 m_1^2 + x_2 m_2^2}.
\eq
The analytic result for this integral is well known:
\bq
\label{one_loop_analytic_solution}
 B\left( t \right) & = &
 \frac{\mu^2}{\sqrt{p_1(t)}} \ln\left( \frac{m_1^2+m_2^2-t + \sqrt{p_1(t)}}{m_1^2+m_2^2-t - \sqrt{p_1(t)}} \right),
 \nonumber \\
 p_1(t) & = & \left[ \left(m_1+m_2\right)^2-t\right]\left[ \left(m_1-m_2\right)^2-t\right].
\eq
The analytical result is easily obtained by direct integration in eq.~(\ref{def_one_loop}).
As an exercise for the extension towards two-loops we derive this result in the following in an alternative way through a differential equation.
The differential equation for $B(t)$ reads
\bq
 \left[ p_1(t) \frac{d}{dt} + p_0(t) \right] B\left( t \right) & = & -2 \mu^2,
 \nonumber \\
 p_0(t) & = & t - m_1^2 - m_2^2.
\eq
A solution of the homogeneous equation is given by
\bq
 \psi\left( t \right) & = & \exp\left(- \int\limits_0^t dv \frac{p_0(v)}{p_1(v)} \right)
 = \frac{m_2^2-m_1^2}{\sqrt{p_1(t)}}.
\eq
The boundary value at $t=0$ is given by
\bq
 B\left(0\right) & = &
 \frac{\mu^2}{m_2^2-m_1^2} \ln \frac{m_2^2}{m_1^2}.
\eq
The solution of the inhomogeneous equation can be written in the form
\bq
\label{one_loop_solution_diff_eq}
 B\left( t \right) & = &
 C \psi\left( t \right) 
 + \psi\left( t \right) \int\limits_0^t dv \frac{\left(-2\mu^2\right)}{p_1(v) \psi\left( v \right)}.
\eq
With $\psi(0)=1$ and $C=B(0)$ we have
\bq
\label{one_loop_homogeneous_solution_diff_eq}
 C \psi\left( t \right) 
 & = &
 \frac{\mu^2}{\sqrt{p_1(t)}} \ln \frac{m_2^2}{m_1^2}.
\eq
The second term in eq.~(\ref{one_loop_solution_diff_eq}) is a special solution of the inhomogeneous equation. 
Direct integration leads to
\bq
\label{one_loop_special_solution_diff_eq}
 \psi\left( t \right) \int\limits_0^t dv \frac{\left(-2\mu^2\right)}{p_1(v) \psi\left( v \right)}
 & = & 
 \frac{\mu^2}{\sqrt{p_1(t)}} \ln\left( \frac{m_1^2 \left( m_1^2+m_2^2-t+\sqrt{p_1(t)}\right)}{m_2^2 \left( m_1^2+m_2^2-t-\sqrt{p_1(t)}\right)} \right).
\eq
Plugging the results of eq.~(\ref{one_loop_homogeneous_solution_diff_eq}) and eq.~(\ref{one_loop_special_solution_diff_eq}) 
into eq.~(\ref{one_loop_solution_diff_eq}) we recover the analytical solution in eq.~(\ref{one_loop_analytic_solution}).
The argument of the logarithm can be re-written as
\bq
 \frac{m_1^2+m_2^2-t + \sqrt{p_1(t)}}{m_1^2+m_2^2-t - \sqrt{p_1(t)}}
 & = &
 \left( \frac{\sqrt{(m_1+m_2)^2-t} + \sqrt{(m_1-m_2)^2-t}}{\sqrt{(m_1+m_2)^2-t} - \sqrt{(m_1-m_2)^2-t}} \right)^2
\eq
and we finally obtain
\bq
\label{final_result_one_loop}
 B\left( t \right) = 
 \frac{2 \mu^2}{\sqrt{( (m_1+m_2)^2-t)( (m_1-m_2)^2-t)}} \ln\left( \frac{\sqrt{(m_1+m_2)^2-t} + \sqrt{(m_1-m_2)^2-t}}{\sqrt{(m_1+m_2)^2-t} - \sqrt{(m_1-m_2)^2-t}} \right).
\eq
We see that the solution consists of an algebraic prefactor $\mu^2/\sqrt{p_1(t)}$ and a transcendental function $\ln(...)$.
We note that the algebraic prefactor can already be obtained from the homogeneous solution in eq.~(\ref{one_loop_homogeneous_solution_diff_eq}).
The point $t=(m_1+m_2)^2$ is called the threshold, the point $t=(m_1-m_2)^2$ is called the pseudo-threshold.
On the principal branch of the logarithm the integral is finite at the pseudo-threshold.
This is easily seen from eq.~(\ref{final_result_one_loop}): At the pseudo-threshold the argument of the logarithm is approaching
one, thus cancelling the square-root singularity of the prefactor.

\section{Definition of the two-loop sunrise integral}
\label{sec:definition}

\begin{figure}
\begin{center}
\begin{picture}(100,100)(0,0)
\Vertex(20,50){2}
\Vertex(80,50){2}
\Line(20,50)(80,50)
\Line(80,50)(100,50)
\Line(0,50)(20,50)
\CArc(50,50)(30,0,180)
\CArc(50,50)(30,180,360)
\Text(105,50)[l]{$p$}
\Text(50,85)[b]{$m_1$}
\Text(50,55)[b]{$m_2$}
\Text(50,25)[b]{$m_3$}
\end{picture}
\end{center}
\caption{
The two-loop sunrise graph.
}
\label{fig_sunrise_graph}
\end{figure}
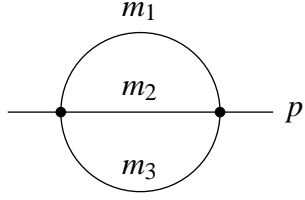

The two-loop integral corresponding to the sunrise graph with arbitrary masses is given 
in $2$-dimensional Minkowski space by
\bq
\label{def_sunrise}
 S\left( p^2 \right)
 & = &
 \mu^2
 \int \frac{d^2k_1}{i \pi} \frac{d^2k_2}{i \pi}
 \frac{1}{\left(-k_1^2+m_1^2\right)\left(-k_2^2+m_2^2\right)\left(-\left(p-k_1-k_2\right)^2+m_3^2\right)}.
\eq
The corresponding sunrise graph is shown in fig.~(\ref{fig_sunrise_graph}). In eq.~(\ref{def_sunrise}) the three internal masses 
are denoted by $m_1$, $m_2$ and $m_3$. 
Without loss of generality we assume that the masses are ordered as
\bq
\label{ordering_masses}
 0 < m_1 \le m_2 \le m_3.
\eq
The arbitrary scale $\mu$ is introduced to make the integral dimensionless.
$p^2$ denotes the momentum squared. This variable plays an important role in our derivation and it is convenient
to introduce the notation
\bq
 t & = & p^2.
\eq
In terms of Feynman parameters the two-loop integral is given by
\bq
\label{def_Feynman_integral}
 S\left( t\right)
 & = & 
 \mu^2
 \int\limits_\sigma \frac{\omega}{{\cal F}} 
\eq
with
\bq
 {\cal F} & = & - x_1 x_2 x_3 t
                + \left( x_1 m_1^2 + x_2 m_2^2 + x_3 m_3^2 \right) {\cal U},
 \nonumber \\
 {\cal U} & = & x_1 x_2 + x_2 x_3 + x_3 x_1.
\eq
The differential two-form $\omega$ is given by
\bq
 \omega & = & x_1 dx_2 \wedge dx_3 + x_2 dx_3 \wedge dx_1 + x_3 dx_1 \wedge dx_2.
\eq
Eq.~(\ref{def_Feynman_integral}) is a projective integral and we can integrate over any surface covering the solid angle of the first octant:
\bq
 \sigma & = & \left\{ \left[ x_1 : x_2 : x_3 \right] \in {\mathbb P}^2 | x_i \ge 0, i=1,2,3 \right\}.
\eq
It will be convenient to introduce the following notation:
We denote the masses related to the pseudo-thresholds by
\bq
\label{def_pseudo_thresholds}
 \mu_1 = m_1+m_2-m_3,
 \;\;\;
 \mu_2 = m_1-m_2+m_3,
 \;\;\;
 \mu_3 = -m_1+m_2+m_3,
\eq
and the mass related to the threshold by
\bq
\label{def_thresholds}
 \mu_4 = m_1+m_2+m_3.
\eq
With the ordering of eq.~(\ref{ordering_masses}) we have
\bq
 \mu_1 \le \mu_2 \le \mu_3 < \mu_4.
\eq
We further introduce the monomial symmetric polynomials $M_{\lambda_1 \lambda_2 \lambda_3}$ in the variables $m_1^2$, $m_2^2$ and $m_3^2$.
These are defined by
\bq
 M_{\lambda_1 \lambda_2 \lambda_3} & = &
 \sum\limits_{\sigma} \left( m_1^2 \right)^{\sigma\left(\lambda_1\right)} \left( m_2^2 \right)^{\sigma\left(\lambda_2\right)} \left( m_3^2 \right)^{\sigma\left(\lambda_3\right)},
\eq
where the sum is over all distinct permutations of $\left(\lambda_1,\lambda_2,\lambda_3\right)$.
A few examples are
\bq
 M_{100} & = & m_1^2 + m_2^2 + m_3^2,
 \nonumber \\
 M_{111} & = & m_1^2 m_2^2 m_3^2,
 \nonumber \\
 M_{210} & = & m_1^4 m_2^2 + m_2^4 m_3^2 + m_3^4 m_1^2 + m_2^4 m_1^2 + m_3^4 m_2^2 + m_1^4 m_3^2.
\eq
We also introduce the abbreviations
\bq
\label{def_delta}
 & & \Delta =  2 M_{110} - M_{200} = \mu_1 \mu_2 \mu_3 \mu_4,
 \nonumber \\
 & &
 \delta_1 = -m_1^2 + m_2^2 + m_3^2,
 \;\;\;\;\;\;
 \delta_2 = m_1^2 - m_2^2 + m_3^2,
 \;\;\;\;\;\;
 \delta_3 = m_1^2 + m_2^2 - m_3^2.
\eq
As shown in \cite{MullerStach:2011ru} the two-loop sunrise integral in $D=2$ dimensions satisfies a second-order differential equation:
\bq
\label{second_order_dgl}
 \left[ p_0(t) \frac{d^2}{d t^2} + p_1(t) \frac{d}{dt} + p_2(t) \right] S\left(t\right) & = & \mu^2 p_3(t),
\eq
where $p_0(t)$, $p_1(t)$, $p_2(t)$ and $p_3(t)$ are polynomials in $t$.
(We follow here the notation of \cite{MullerStach:2011ru} and denote the coefficient of $d^j/dt^j$ by $p_{2-j}(t)$.)
The polynomial $p_0(t)$ is given by
\bq
\label{def_p0}
 p_0(t) & = &
  t \left( t - \mu_1^2 \right)
    \left( t - \mu_2^2 \right)
    \left( t - \mu_3^2 \right)
    \left( t - \mu_4^2 \right)
    \left( 3 t^2 - 2 M_{100} t + \Delta \right).
\eq
The zeros of the polynomial $p_0(t)$ correspond to the singular points of the differential equation.
The regular singular points at $t=0$, at the pseudo-thresholds $\mu_1$, $\mu_2$ and $\mu_3$ as well as the threshold $\mu_4$ are expected.
In addition the roots of the quadratic polynomial $3t^2-2M_{100}t+\Delta$ define two regular singular points.
We will comment on these points after eq.~(\ref{expr_wronski}).
The polynomials $p_1(t)$ and $p_2(t)$ read
\bq
 p_1(t) & = & 
  9 t^6
  - 32 M_{100} t^5
  + \left( 37 M_{200} + 70 M_{110} \right) t^4
  - \left( 8 M_{300} + 56 M_{210} + 144 M_{111} \right) t^3
 \\
 & &
  - \left( 13 M_{400} - 36 M_{310} + 46 M_{220} - 124 M_{211} \right) t^2
 \nonumber \\
 & &
  - \left( -8 M_{500} + 24 M_{410} - 16 M_{320} - 96 M_{311} + 144 M_{221} \right) t
 \nonumber \\
 & &
  - \left( M_{600} - 6 M_{510} + 15 M_{420} - 20 M_{330} + 18 M_{411} - 12 M_{321} - 6 M_{222} \right),
 \nonumber \\
 p_2(t) & = &
  3 t^5
  - 7 M_{100} t^4
  + \left( 2 M_{200} + 16 M_{110} \right) t^3
  + \left( 6 M_{300} - 14 M_{210} \right) t^2
 \nonumber \\
 & &
  - \left( 5 M_{400} - 8 M_{310} + 6 M_{220} - 8 M_{211} \right) t
  + \left( M_{500} - 3 M_{410} + 2 M_{320} + 8 M_{311} - 10 M_{221} \right).
 \nonumber
\eq
In contrast to the polynomial $p_0(t)$, which is given in factorised form in eq.~(\ref{def_p0}),
we do not know a factorisation for $p_1(t)$ or $p_2(t)$. However, we can write both polynomials as a sum of two terms, one proportional
to the quadratic polynomial $3t^2-2M_{100}t+\Delta$ and the second proportional to the derivative $6t-2M_{100}$ of the quadratic polynomial:
\bq
\lefteqn{
 p_1(t) = 
 \frac{d}{dt} p_0(t) - 2 p_0(t) \frac{d}{dt} \ln\left(3t^2-2M_{100}t+\Delta \right),
} & &
 \\
\lefteqn{
 p_2(t) = 
 - \frac{1}{4 t} p_0(t) \frac{d}{dt} \ln\left(3t^2-2M_{100}t+\Delta \right)
} & &
 \nonumber \\
 & &
 + \frac{1}{2} \left(3t^2-2M_{100}t+\Delta \right) \left( 3 t^3 - 7 M_{100} t^2 + \left(5M_{200}+6M_{110}\right) t - M_{300} + M_{210} - 18 M_{111} \right).
 \nonumber 
\eq
The polynomial $p_3(t)$ appearing in the inhomogeneous part is given by
\bq
 p_3(t) & = & 
 -2 \left( 3 t^2 - 2 M_{100} t + \Delta \right)^2
 \\
 & & 
 + 2 c\left(t,m_1,m_2,m_3\right)  \ln \frac{m_1^2}{\mu^2}
 + 2 c\left(t,m_2,m_3,m_1\right)  \ln \frac{m_2^2}{\mu^2}
 + 2 c\left(t,m_3,m_1,m_2\right)  \ln \frac{m_3^2}{\mu^2},
 \nonumber 
\eq
with
\bq
\lefteqn{
c\left(t,m_1,m_2,m_3\right) = } & &
 \nonumber \\
 & &
 \left( -2 m_1^2 + m_2^2 + m_3^2 \right) t^3
 + \left( 6 m_1^4 - 3 m_2^4 - 3 m_3^4 - 7 m_1^2 m_2^2 - 7 m_1^2 m_3^2 + 14 m_2^2 m_3^2 \right) t^2
 \nonumber \\
 & &
 + \left( -6 m_1^6 + 3 m_2^6 + 3 m_3^6 + 11 m_1^4 m_2^2 + 11 m_1^4 m_3^2 - 8 m_1^2 m_2^4 - 8 m_1^2 m_3^4 - 3 m_2^4 m_3^2 - 3 m_2^2 m_3^4 \right) t
 \nonumber \\
 & & 
 + \left( 2 m_1^8 - m_2^8 - m_3^8 - 5 m_1^6 m_2^2 - 5 m_1^6 m_3^2 + m_1^2 m_2^6 + m_1^2 m_3^6 + 4 m_2^6 m_3^2 + 4 m_2^2 m_3^6 
 \right. \nonumber \\
 & & \left.
        + 3 m_1^4 m_2^4 + 3 m_1^4 m_3^4 - 6 m_2^4 m_3^4 
        + 2 m_1^4 m_2^2 m_3^2 - m_1^2 m_2^4 m_3^2 - m_1^2 m_2^2 m_3^4 \right).
\eq
We remark that the inhomogeneous second-order differential equation in eq.~(\ref{second_order_dgl})
implies that $S(t)$ is a solution of a factorised homogeneous third-order differential equation:
\bq
 \left[ p_3(t) \frac{d}{dt} - \dot{p}_3(t) \right]
 \left[ p_0(t) \frac{d^2}{d t^2} + p_1(t) \frac{d}{dt} + p_2(t) \right] S\left(t\right) & = & 0,
\eq
with 
\bq
 \dot{p}_3(t) & = &
 \frac{d}{dt} p_3(t).
\eq

\section{The solution of the differential equation}
\label{sec:solution}

We will solve the second-order differential equation in eq.~(\ref{second_order_dgl})
by first considering the corresponding homogeneous equation. We denote the two independent solutions of the
homogeneous equation by $\psi_1(t)$ and $\psi_2(t)$.
The solution of the inhomogeneous differential equation is then obtained through the method of the variation of the constants.
The full solution can be written as
\bq
\label{inhomogeneous_solution}
 S\left(t\right)
 & = & 
 C_1 \psi_1(t) + C_2 \psi_2(t)
 + \mu^2 \int\limits_{0}^{t} dt_1 \frac{p_3(t_1)}{p_0(t_1) W(t_1)} \left[ - \psi_1(t) \psi_2(t_1) + \psi_2(t) \psi_1(t_1) \right],
\eq
where $W(t)$ denotes the Wronski determinant.
In the next sub-sections we will first determine the homogeneous solutions $\psi_1(t)$ and $\psi_2(t)$, then fix the constants
$C_1$ and $C_2$ from the boundary values and finally consider a special solution of the inhomogeneous equation, given
by the third term in eq.~(\ref{inhomogeneous_solution}).

The integral $S(t)$ is real in the Euclidean region defined by $t<0$ and $m_i>0$. The analytical continuation to $t>0$ is provided
by Feynman's $i0$-prescription, where we substitute $t \rightarrow t + i0$. 
The symbol $+i0$ denotes an infinitesimal positive imaginary part.
We will often encounter expressions of the form $\sqrt{m^2-t}$. We have
\bq
 \sqrt{m^2-t-i0}
 \in
 \left\{ \begin{array}{rll}
 {\mathbb R}^+ & \mbox{for} & t<m^2, \\
 -i{\mathbb R}^+ & \mbox{for} & t>m^2. \\
 \end{array} \right.
\eq
We can factorise this expression as
\bq
 \sqrt{m^2-t-i0}
 & = &
 \sqrt{\left(m-\sqrt{t+i0}\right)\left(m+\sqrt{t+i0}\right)}.
\eq
Note that
\bq
 \sqrt{t+i0} = i \sqrt{-t-i0}
 & = &
 \left\{ \begin{array}{cll}
 i \left(\sqrt{-t}-i0\right)  & \mbox{for} & t<0, \\
 \sqrt{t} + i0 & \mbox{for} & t>0. \\
 \end{array} \right.
\eq
In the following we will suppress the $i0$-part and write simply
\bq
 \sqrt{m^2-t}
 & = &
 \sqrt{\left(m-\sqrt{t}\right)\left(m+\sqrt{t}\right)}.
\eq

\subsection{The homogeneous solutions}
\label{sec:homogeneous}

In this section we consider the solution of the homogeneous differential equation
\bq
\label{homogeneous_eq}
 \left[ p_0(t) \frac{d^2}{d t^2} + p_1(t) \frac{d}{dt} + p_2(t) \right] \psi\left(t\right) & = & 0.
\eq
We will denote the two linear independent solutions of eq.~(\ref{homogeneous_eq}) by $\psi_1(t)$ and $\psi_2(t)$.
We recall that the Wronski determinant is defined by
\bq
 W(t) & = & \psi_1(t) \frac{d}{dt} \psi_2(t) - \psi_2(t) \frac{d}{dt} \psi_1(t)
\eq
and satisfies
\bq
 \frac{d}{dt} W(t) & = &
 - \frac{p_1(t)}{p_0(t)} W(t).
\eq
This equation can easily be integrated and yields
\bq
\label{expr_wronski}
 W(t) & = & 
 c' \frac{\left( 3 t^2 - 2 M_{100}t + \Delta \right)}{t \left( t - \mu_1^2 \right) \left( t - \mu_2^2 \right) \left( t - \mu_3^2 \right) \left( t - \mu_4^2 \right)}.
\eq
The integration constant $c'$ will be fixed later.
The Wronski determinant vanishes at the regular singular points defined by the roots of the equation $3t^2-2M_{100}t+\Delta=0$.

In order to present the solutions of the homogeneous equation we introduce the notation
\bq
\label{def_t_values}
 x_1 = \frac{\left(m_1-m_2\right)^2}{\mu^2},
\;\;\;
 x_2 = \frac{\left(m_3-\sqrt{t}\right)^2}{\mu^2},
\;\;\;
 x_3 = \frac{\left(m_3+\sqrt{t}\right)^2}{\mu^2},
\;\;\;
 x_4 = \frac{\left(m_1+m_2\right)^2}{\mu^2}.
\eq
For $t<0$ the quantities $x_2$ and $x_3$ are complex. For $t>0$ the four quantities are real.
If in addition to $t>0$ we assume $m_3^2<(m_1+m_2)^2$ and that $t$ is small they are ordered as
\bq
 0 \le x_1 < x_2 < x_3 < x_4.
\eq
The equation
\bq
\label{elliptic_curve}
 y^2 & = & \left(x-x_1\right)\left(x-x_2\right)\left(x-x_3\right)\left(x-x_4\right)
\eq
defines an elliptic curve. The holomorphic one-form associated to this elliptic curve is $dx/y$ and the 
periods of the elliptic curve are given by
\bq
\label{solution_hom}
 \psi_1\left(t\right) & = & 
 2 \int\limits_{x_2}^{x_3} \frac{dx}{y} 
 \;\; = \;\;
 \frac{4}{\sqrt{\left(x_3-x_1\right)\left(x_4-x_2\right)}} \; K\left(\sqrt{ \frac{\left(x_3-x_2\right)\left(x_4-x_1\right)}{\left(x_3-x_1\right)\left(x_4-x_2\right)}}\right),
 \nonumber \\
 \psi_2\left(t\right) & = & 
 2 \int\limits^{x_3}_{x_4} \frac{dx}{y} 
 \;\; = \;\;
 \frac{4 i}{\sqrt{\left(x_3-x_1\right)\left(x_4-x_2\right)}} \; K\left(\sqrt{ \frac{\left(x_2-x_1\right)\left(x_4-x_3\right)}{\left(x_3-x_1\right)\left(x_4-x_2\right)}}\right).
\eq
$K(k)$ denotes the complete elliptic integral of the first kind
\bq
 K(k)
 & = &
 \int\limits_0^1 \frac{dt}{\sqrt{\left(1-t^2\right)\left(1-k^2t^2\right)}}.
\eq
The modulus $k(t)$ of $\psi_1(t)$ is given by
\bq
\label{def_modulus}
 k\left(t\right)
 & = &
 \sqrt{ \frac{\left(x_3-x_2\right)\left(x_4-x_1\right)}{\left(x_3-x_1\right)\left(x_4-x_2\right)}}
 =
 \sqrt{ \frac{16 m_1 m_2 m_3 \sqrt{t}}{\left(\mu_1+\sqrt{t}\right)\left(\mu_2+\sqrt{t}\right)\left(\mu_3+\sqrt{t}\right)\left(\mu_4-\sqrt{t}\right)} }.
\eq
We recall that the pseudo-thresholds $\mu_1$, $\mu_2$, $\mu_3$ and the threshold $\mu_4$ have been defined in eq.~(\ref{def_pseudo_thresholds})
and in eq.~(\ref{def_thresholds}), respectively.
The complementary modulus is given by
\bq
\label{def_complementary_modulus}
 k'\left(t\right)
 & = &
 \sqrt{ \frac{\left(x_2-x_1\right)\left(x_4-x_3\right)}{\left(x_3-x_1\right)\left(x_4-x_2\right)}}
 =
 \sqrt{ \frac{\left(\mu_1-\sqrt{t}\right)\left(\mu_2-\sqrt{t}\right)\left(\mu_3-\sqrt{t}\right)\left(\mu_4+\sqrt{t}\right)}
             {\left(\mu_1+\sqrt{t}\right)\left(\mu_2+\sqrt{t}\right)\left(\mu_3+\sqrt{t}\right)\left(\mu_4-\sqrt{t}\right)} }.
\eq
Note that we have the relation
\bq
 k(t)^2 + k'(t)^2 & = & 1.
\eq
We would like to remark that although we started in eq.~(\ref{elliptic_curve}) from an equation 
which is not symmetric with respect to the permutation of the masses $m_1$, $m_2$ and $m_3$,
the functions $\psi_1(t)$ and $\psi_2(t)$ are symmetric in the masses.
We further remark that the elliptic curve defined in eq.~(\ref{elliptic_curve}) varies with $t$, and so do the periods
$\psi_1(t)$ and $\psi_2(t)$.

The main result of this sub-section is the following:
The periods $\psi_1(t)$ and $\psi_2(t)$ defined in eq.~(\ref{solution_hom}) are two
independent solutions of the homogeneous differential equation.
This is easily verified by inserting the functions $\psi_1(t)$ and $\psi_2(t)$ into eq.~(\ref{homogeneous_eq}).
The solutions have been found by generalising the corresponding solutions for the equal mass case in ref.~\cite{Laporta:2004rb} to the unequal
mass case.
The appropriate values for the variables $x_1$, $x_2$, $x_3$ and $x_4$ for the unequal mass case have been suggested in \cite{Bauberger:1994nk,Bauberger:1994by}.
The representation in eq.~(\ref{solution_hom}) of the homogeneous solutions is not unique. To see this in an example we set
\bq
\label{def_X_i}
 X_1(t) & = & \left(x_3-x_2\right)\left(x_4-x_1\right) =
 16 m_1 m_2 m_3 \sqrt{t} / \mu^4,
 \nonumber \\
 X_2(t) & = & \left(x_2-x_1\right)\left(x_4-x_3\right) =
 \left(\mu_1-\sqrt{t}\right)\left(\mu_2-\sqrt{t}\right)\left(\mu_3-\sqrt{t}\right)\left(\mu_4+\sqrt{t}\right) / \mu^4,
 \nonumber \\
 X_3(t) & = & \left(x_3-x_1\right)\left(x_4-x_2\right) =
 \left(\mu_1+\sqrt{t}\right)\left(\mu_2+\sqrt{t}\right)\left(\mu_3+\sqrt{t}\right)\left(\mu_4-\sqrt{t}\right) / \mu^4.
\eq
We have
\bq
 X_1 + X_2 & = & X_3.
\eq
Then any of the six functions
\bq
 & &
 \psi_1 = \frac{4}{\sqrt{X_3}} \;\; K\left(\sqrt{\frac{X_1}{X_3}}\right),
 \;\;\;\;\;
 \psi_3 = \frac{4}{\sqrt{X_1}} \;\; K\left(\sqrt{\frac{X_3}{X_1}}\right),
 \;\;\;\;\;
 \psi_5 = \frac{4}{\sqrt{X_2}} \;\; K\left(i\sqrt{\frac{X_1}{X_2}}\right),
 \nonumber \\
 & &
 \psi_2 = \frac{4i}{\sqrt{X_3}} \;\; K\left(\sqrt{\frac{X_2}{X_3}}\right),
 \;\;\;\;\;
 \psi_4 = \frac{4i}{\sqrt{X_2}} \;\; K\left(\sqrt{\frac{X_3}{X_2}}\right),
 \;\;\;\;\;
 \psi_6 = \frac{4i}{\sqrt{X_1}} \;\; K\left(i\sqrt{\frac{X_2}{X_1}}\right),
\eq
is a solution of the homogeneous differential equation.
Of course, an ordinary second-order differential equation has only two independent solutions, therefore
there must be relations among these functions.
Indeed one finds \cite{fettis:1970:rmr}
\bq
 \psi_3= \psi_1- s \psi_2,
 \;\;\;
 \psi_4 = -s \psi_1+\psi_2,
 \;\;\;
 \psi_5 = \psi_1,
 \;\;\;
 \psi_6 = \psi_2,
\eq
where $s=\mathrm{sign}(\mathrm{Im}(X_1/X_3))$.
We have chosen the functions $\psi_1$ and $\psi_2$ as a basis. With this choice both the modulus $k$ and the complementary modulus $k'$ are real
and in the interval $[0,1]$ (for $m_3^2<(m_1+m_2)^2$ and small positive $t$).

The function $\psi_1$ is regular at $t=0$ and we find the value
\bq
 \psi_1\left(0\right) & = & \frac{2 \pi \mu^2}{\sqrt{\Delta}}.
\eq
On the other hand, the function $\psi_2$ exhibits a logarithmic singularity at the origin. We have
\bq
 \lim\limits_{t\rightarrow 0} \left( t \frac{d}{dt} \psi_2\left(t\right) \right)
 & = &
 - \frac{i \mu^2}{\sqrt{\Delta}}.
\eq
From the solution in eq.~(\ref{solution_hom}) we find that the constant $c'$ appearing in the Wronski determinant is given by
\bq
 c' & = & - 2 \pi i \left( \mu^2 \right)^2.
\eq
In simplifying the expression for the Wronski determinant the Legendre relation
\bq
 K\left(k\right) E\left(k'\right) + E\left(k\right) K\left(k'\right) - K\left(k\right) K\left(k'\right) & = & \frac{\pi}{2}
\eq
is useful. $k$ and $k'$ are the modulus and the complementary modulus defined in eq.~(\ref{def_modulus}) and eq.~(\ref{def_complementary_modulus}), respectively.
$E(x)$ denotes the complete elliptic integral of the second kind. (The definition will be given in eq.~(\ref{def_elliptic_integral_second_kind}).)
 
\subsection{The boundary values}
\label{sec:boundary}

In this section we determine the two constants $C_1$ and $C_2$ from the boundary value at $t=0$.
We have already seen that the homogeneous solution $\psi_2$ has a logarithmic singularity at $t=0$.
On the other hand the sunrise integral is regular at $t=0$.
Therefore 
\bq
 C_2 & = & 0.
\eq
The constant $C_1$ is given by
\bq
 C_1 & = & \frac{\sqrt{\Delta}}{2 \pi \mu^2} S_0,
\eq
where $S_0$ is the sunrise integral at $t=0$:
\bq
 S_0
 & = & 
 \mu^2
 \int\limits_\sigma \frac{\omega}{\left( x_1 m_1^2 + x_2 m_2^2 + x_3 m_3^2 \right) {\cal U}}.
\eq
We can integrate over any surface covering the solid angle of the first octant and it is convenient to do the integration in the plane
\bq
 x_1 m_1^2 + x_2 m_2^2 + x_3 m_3^2 & = & M^2,
\eq
where $M^2$ is an arbitrary constant. The change of variables
\bq
 x_1' = \frac{m_1^2}{M^2} x_1,
 \;\;\;
 x_2' = \frac{m_2^2}{M^2} x_2,
 \;\;\;
 x_3' = \frac{m_3^2}{M^2} x_3
\eq
transforms the integral into
\bq
 S_0
 & = & 
 \mu^2
 \int\limits_\sigma \frac{\omega}{m_1^2 x_2 x_3 + m_2^2 x_3 x_1 + m_3^2 x_1 x_2}.
\eq
This is nothing else than the Feynman parameter integral for the one-loop three-point function in four dimensions
with massless internal lines and three external masses \cite{Davydychev:1995mq}.
With $p_1+p_2+p_3=0$ and 
\bq
 T\left( 4, p_1^2, p_2^2, p_3^2, \mu^2 \right)
 & = &
 \mu^2
 \int \frac{d^4k}{i \pi^2} 
 \frac{1}{\left(-k^2\right)\left(-(k-p_1)^2\right)\left(-\left(k-p_1-p_2\right)^2\right)}
 \nonumber \\
 & = &
 \mu^2
 \int\limits_\sigma \frac{\omega}{(-p_1^2) x_2 x_3 + (-p_2^2) x_3 x_1 + (-p_3^2) x_1 x_2}
\eq
we have
\bq
 S_0 & = &
 T\left( 4, -m_1^2, -m_2^2, -m_3^2, \mu^2 \right).
\eq
In a neighbourhood of the equal mass point we have $\Delta>0$ and the integral $S_0$ is expressed in the region
$\Delta > 0$  by \cite{Ussyukina:1993jd,Lu:1992ny,Bern:1997ka}
\bq
 S_0 = 
 \frac{2\mu^2}{\sqrt{\Delta}} 
 \left[ \mbox{Cl}_2\left( \alpha_1 \right) + \mbox{Cl}_2\left( \alpha_2 \right) + \mbox{Cl}_2\left( \alpha_3 \right) \right].
\eq
The Clausen function $\mbox{Cl}_2(x)$ is given in terms of dilogarithms by
\bq
\label{defclausen}
 \mbox{Cl}_2(x) & = & \frac{1}{2i} \left[ \mbox{Li}_2\left(e^{ix}\right) - \mbox{Li}_2\left(e^{-ix}\right) \right].
\eq
The arguments $\alpha_i$ of the Clausen functions are defined by
\bq
 \alpha_i & = & 2 \arctan \left( \frac{\sqrt{\Delta}}{\delta_i}\right),
 \;\;\;\;\;\;
 i \in \{1,2,3\}.
\eq
We recall that $\Delta$ and the quantities $\delta_i$ have been defined in eq.~(\ref{def_delta}).
Let us summarise what we obtained so far: The homogeneous solution with the correct boundary value at $t=0$ is given by
\bq
\lefteqn{
 C_1 \psi_1(t) + C_2 \psi_2(t) = 
} & &
 \\
 & &
 \frac{4 \mu^2 K\left(k(t)\right)}{\pi \sqrt{\left(\mu_1+\sqrt{t}\right)\left(\mu_2+\sqrt{t}\right)\left(\mu_3+\sqrt{t}\right)\left(\mu_4-\sqrt{t}\right)}}
 \left[ \mbox{Cl}_2\left( \alpha_1 \right) + \mbox{Cl}_2\left( \alpha_2 \right) + \mbox{Cl}_2\left( \alpha_3 \right) \right].
 \nonumber
\eq

\subsection{The inhomogeneous solutions}
\label{sec:inhomogeneous}

In this section we consider a special solution of the inhomogeneous equation.
Variation of the constants leads to
\bq
 S_{\mathrm{special}}\left(t\right)
 & = & 
 \mu^2 \int\limits_{0}^{t} dt_1 \frac{p_3(t_1)}{p_0(t_1) W(t_1)} \left[ - \psi_1(t) \psi_2(t_1) + \psi_2(t) \psi_1(t_1) \right].
\eq
Note that
\bq
\label{eq_inhom_factor}
 \mu^2 \frac{p_3(t)}{p_0(t) W(t)} & = & 
 \frac{1}{i \pi \mu^2}
 - \frac{1}{i \pi \mu^2}
   \frac{1}{\left( 3 t^2 - 2 M_{100}t + \Delta \right)^{2}}
 \\
 & & 
 \times 
   \left[ c\left(t,m_1,m_2,m_3\right)  \ln \frac{m_1^2}{\mu^2}
 + c\left(t,m_2,m_3,m_1\right)  \ln \frac{m_2^2}{\mu^2}
 + c\left(t,m_3,m_1,m_2\right)  \ln \frac{m_3^2}{\mu^2} \right].
 \nonumber
\eq
The function $c\left(t,m_1,m_2,m_3\right)$ is a polynomial of degree $3$ in the variable $t$.
Let us denote the roots of the quadratic equation
\bq
 3 t^2 - 2 M_{100}t + \Delta & = & 0
\eq
by $t_+$ and $t_-$. We have
\bq
 t_\pm & = & \frac{1}{3} \left( M_{100} \pm 2 \sqrt{M_{200}-M_{110}} \right).
\eq 
Partial fraction decomposition leads to
\bq
 \mu^2 \frac{p_3(t)}{p_0(t) W(t)} & = & 
 \frac{1}{i \pi \mu^2}
 \left[
 1 
 + \frac{a_2\left(t_+,m_1,m_2,m_3\right)}{\left(t-t_+\right)^2} 
 + \frac{a_2\left(t_-,m_1,m_2,m_3\right)}{\left(t-t_-\right)^2} 
 \right.
 \nonumber \\
 & &
 \left.
 + \frac{a_1\left(t_+,m_1,m_2,m_3\right)}{\left(t_+-t_-\right)\left(t-t_+\right)} 
 - \frac{a_1\left(t_-,m_1,m_2,m_3\right)}{\left(t_+-t_-\right)\left(t-t_-\right)} 
 \right]
\eq
with
\bq
 a_i\left(t,m_1,m_2,m_3\right) = 
          c_i\left(t,m_1,m_2,m_3\right) \ln \frac{m_1^2}{\mu^2} 
        + c_i\left(t,m_2,m_3,m_1\right) \ln \frac{m_2^2}{\mu^2} 
        + c_i\left(t,m_3,m_1,m_2\right) \ln \frac{m_3^2}{\mu^2} 
 \nonumber 
\eq
and
\bq
\lefteqn{
 c_1\left(t,m_1,m_2,m_3\right) = 
 \frac{1}{9} \left[ -2 m_1^4 + m_2^4 + m_3^4 +m_1^2 m_2^2 + m_1^2 m_3^2 - 2 m_2^2 m_3^2 +t \left( 2 m_1^2 - m_2^2 - m_3^2 \right) \right],
} & &
 \nonumber \\
\lefteqn{
 c_2\left(t,m_1,m_2,m_3\right) = 
 \frac{1}{9 \left( M_{200} - M_{110} \right)}
 \left[ \Delta (2m_1^4-m_2^4-m_3^4-2m_1^2m_2^2-2m_1^2m_3^2+4m_2^2m_3^2)
 \right.
} & &
 \nonumber \\
 & &
 \left.
        +t( 2m_1^6-m_2^6-m_3^6 + 9m_1^2m_2^4+9m_1^2m_3^4 - 6m_1^4m_2^2-6m_1^4m_3^2 -3m_2^4m_3^2-3m_2^2m_3^4 )
 \right].
 \hspace*{10mm}
\eq
We therefore have
\bq
\label{special_solution_1}
\lefteqn{
 S_{\mathrm{special}}\left(t\right)
 =  
 \frac{1}{i \pi \mu^2}
 \int\limits_{0}^{t} dt_1
 \left[ \psi_2\left(t\right) \psi_1\left(t_1\right) - \psi_1\left(t\right) \psi_2\left(t_1\right) \right]
} & &
 \\
 & &
 \left[
 1 
 + \frac{a_2\left(t_+,m_1,m_2,m_3\right)}{\left(t_1-t_+\right)^2} 
 + \frac{a_2\left(t_-,m_1,m_2,m_3\right)}{\left(t_1-t_-\right)^2} 
 + \frac{a_1\left(t_+,m_1,m_2,m_3\right)}{\left(t_+-t_-\right)\left(t_1-t_+\right)} 
 - \frac{a_1\left(t_-,m_1,m_2,m_3\right)}{\left(t_+-t_-\right)\left(t_1-t_-\right)} 
 \right].
 \nonumber 
\eq
We expect the poles at $t_\pm$ to be non-physical and spurious. They can be eliminated as follows:
We recall that the functions $\psi_1(t)$ and $\psi_2(t)$ have been defined as
\bq
 \psi_1\left(t\right) = \frac{4}{\sqrt{X_3\left(t\right)}} K\left(k\left(t\right)\right),
 & &
 \psi_2\left(t\right) = \frac{4i}{\sqrt{X_3\left(t\right)}} K\left(k'\left(t\right)\right).
\eq
Let us define two associated functions $\phi_1(t)$ and $\phi_2(t)$ by
\bq
 \phi_1\left(t\right) = \frac{4}{\sqrt{X_3\left(t\right)}} \left[ K\left(k\left(t\right)\right) - E\left(k\left(t\right)\right) \right],
 & & 
 \phi_2\left(t\right) = \frac{4 i}{\sqrt{X_3\left(t\right)}} E\left(k'\left(t\right)\right),
\eq
where $E(x)$ denotes the complete elliptic integral of the second kind:
\bq
\label{def_elliptic_integral_second_kind}
 E(x) & = &
 \int\limits_0^1 dt \frac{\sqrt{1-x^2 t^2}}{\sqrt{1-t^2}}.
\eq
The four functions $\psi_1(t)$, $\psi_2(t)$, $\phi_1(t)$ and $\phi_2(t)$ are the entries of the period matrix $P$ of the elliptic curve in eq.~(\ref{elliptic_curve}):
\bq
 P & = &
 \left(\begin{array}{cc}
 \psi_1(t) & \psi_2(t) \\
 \phi_1(t) & \phi_2(t) \\
 \end{array} \right).
\eq
There is an algebraic relation between the four functions $\psi_1$, $\psi_2$, $\phi_1$ and $\phi_2$.
From the Legendre relation it follows that
\bq
 \psi_1(t) \phi_2(t) - \psi_2(t) \phi_1(t) & = & \frac{8 \pi i}{X_3(t)}.
\eq
In order to eliminate the double poles $1/(t_1-t_\pm)^2$ we start from the identity
\bq
\label{ibp_1}
 \frac{1}{\left(t_1-t_\pm\right)^2}
 & = &
 - \frac{d}{dt_1} \left( \frac{1}{t_1-t_\pm} + \frac{1}{t_\pm} \right)
\eq
and use integration-by-parts. The constant term on the right-hand side of eq.~(\ref{ibp_1}) ensures that boundary terms vanish.
The derivatives of $\psi_1$, $\psi_2$, $\phi_1$ and $\phi_2$ are given by
\bq
\label{derivative_psi_phi}
 \frac{d}{dt} \psi_i & = &  
 - \frac{1}{2} \psi_i \frac{d}{dt} \ln X_2 + \frac{1}{2} \phi_i \frac{d}{dt} \ln \frac{X_2}{X_1},
 \nonumber \\
 \frac{d}{dt} \phi_i & = & 
 - \frac{1}{2} \psi_i \frac{d}{dt} \ln \frac{X_2}{X_3}
 + \frac{1}{2} \phi_i \frac{d}{dt} \ln \frac{X_2}{X_3^2}.
\eq
We recall from the definition of $X_i(t)$ in eq.~(\ref{def_X_i}) that these functions factorise in the variable $\sqrt{t}$.
As a consequence,
\bq
 \frac{d}{dt} \ln X_i(t)
\eq
has a partial fraction decomposition in $\sqrt{t}$.
We denote
\bq
 \eta_1\left(t_1\right) 
 & = &
 \psi_2\left(t\right) \psi_1\left(t_1\right) - \psi_1\left(t\right) \psi_2\left(t_1\right),
 \nonumber \\
 \eta_2\left(t_1\right) 
 & = &
 \psi_2\left(t\right) \phi_1\left(t_1\right) - \psi_1\left(t\right) \phi_2\left(t_1\right).
\eq
Using eq.~(\ref{ibp_1}) for integration-by-parts in combination with eq.~(\ref{derivative_psi_phi}) one finds that this procedure 
does not only eliminate the double poles $1/(t_1-t_\pm)^2$, but also that the coefficients of the single poles $1/(t_1-t_\pm)$ are zero.
We obtain
\bq
\label{special_solution_2}
 S_{\mathrm{special}}\left(t\right)
 & = & 
 \frac{1}{i \pi \mu^2}
 \int\limits_{0}^{t} dt_1
 \left\{
 \eta_1\left(t_1\right) 
 - \frac{b_1\left(m_1,m_2,m_3\right) t_1 - b_0\left(m_1,m_2,m_3\right)}{3 \mu^4 X_2\left(t_1\right)} \left[ \eta_2\left(t_1\right) - \eta_1\left(t_1\right) \right]
 \right\},
 \nonumber \\
\eq
with
\bq
 b_i\left(m_1,m_2,m_3\right) = 
          d_i\left(m_1,m_2,m_3\right) \ln \frac{m_1^2}{\mu^2} 
        + d_i\left(m_2,m_3,m_1\right) \ln \frac{m_2^2}{\mu^2} 
        + d_i\left(m_3,m_1,m_2\right) \ln \frac{m_3^2}{\mu^2} 
\eq
and
\bq
 d_1\left(m_1,m_2,m_3\right) & = & 2 m_1^2 - m_2^2 - m_3^2 ,
 \nonumber \\
 d_0\left(m_1,m_2,m_3\right) & = & 2 m_1^4 - m_2^4 - m_3^4 -m_1^2 m_2^2 - m_1^2 m_3^2 + 2 m_2^2 m_3^2.
\eq
Note that in the equal mass case $m_1=m_2=m_3$ the terms involving $\eta_2(t_1)-\eta_1(t_1)$ vanish, due to the fact
that
\bq
 d_i\left(m,m,m\right) & = & 0.
\eq

\subsection{The full result}
\label{sec:result}

For the convenience of the reader we put here all pieces together and recall the relevant definitions.
The complete result for the two-loop sunrise integral with arbitrary masses in two space-time dimensions is given by
\bq
\lefteqn{
 S\left(t\right)
 =  
 \frac{1}{\pi}
 \left[ \mbox{Cl}_2\left( \alpha_1 \right) + \mbox{Cl}_2\left( \alpha_2 \right) + \mbox{Cl}_2\left( \alpha_3 \right) \right]
 \psi_1\left(t\right)
} & & \nonumber \\
& &
 +
 \frac{1}{i \pi \mu^2}
 \int\limits_{0}^{t} dt_1
 \left\{
 \eta_1\left(t_1\right) 
 - \frac{b_1\left(m_1,m_2,m_3\right) t_1 - b_0\left(m_1,m_2,m_3\right)}{3 \mu^4 X_2\left(t_1\right)} \left[ \eta_2\left(t_1\right) - \eta_1\left(t_1\right) \right]
 \right\},
\eq
where
\bq
 \eta_1\left(t_1\right) 
 & = &
 \psi_2\left(t\right) \psi_1\left(t_1\right) - \psi_1\left(t\right) \psi_2\left(t_1\right),
 \nonumber \\
 \eta_2\left(t_1\right) 
 & = &
 \psi_2\left(t\right) \phi_1\left(t_1\right) - \psi_1\left(t\right) \phi_2\left(t_1\right).
\eq
The functions $\psi_1$, $\psi_2$, $\phi_1$ and $\phi_2$ are given by
\bq
 \psi_1\left(t\right) = \frac{4}{\sqrt{X_3\left(t\right)}} K\left(k\left(t\right)\right),
 & &
 \psi_2\left(t\right) = \frac{4i}{\sqrt{X_3\left(t\right)}} K\left(k'\left(t\right)\right),
 \nonumber \\
 \phi_1\left(t\right) = \frac{4}{\sqrt{X_3\left(t\right)}} \left[ K\left(k\left(t\right)\right) - E\left(k\left(t\right)\right) \right],
 & & 
 \phi_2\left(t\right) = \frac{4 i}{\sqrt{X_3\left(t\right)}} E\left(k'\left(t\right)\right).
\eq
The modulus $k(t)$ and the complementary modulus $k'(t)$ are defined by
\bq
 k\left(t\right) = \sqrt{\frac{X_1\left(t\right)}{X_3\left(t\right)}},
 & &
 k'\left(t\right) = \sqrt{\frac{X_2\left(t\right)}{X_3\left(t\right)}}.
\eq
The functions $X_1(t)$, $X_2(t)$ and $X_3(t)$ are defined by
\bq
 X_1(t) & = & 
 16 m_1 m_2 m_3 \sqrt{t} / \mu^4,
 \nonumber \\
 X_2(t) & = & 
 \left(\mu_1-\sqrt{t}\right)\left(\mu_2-\sqrt{t}\right)\left(\mu_3-\sqrt{t}\right)\left(\mu_4+\sqrt{t}\right) / \mu^4,
 \nonumber \\
 X_3(t) & = & 
 \left(\mu_1+\sqrt{t}\right)\left(\mu_2+\sqrt{t}\right)\left(\mu_3+\sqrt{t}\right)\left(\mu_4-\sqrt{t}\right) / \mu^4.
\eq
$\mu_1$, $\mu_2$, $\mu_3$ and $\mu_4$ are defined by
\bq
 \mu_1 = m_1+m_2-m_3,
 \;\;\;
 \mu_2 = m_1-m_2+m_3,
 \;\;\;
 \mu_3 = -m_1+m_2+m_3,
 \;\;\;
 \mu_4 = m_1+m_2+m_3.
\eq
The angles $\alpha_1$, $\alpha_2$ and $\alpha_3$ are defined by
\bq
 \alpha_i & = & 2 \arctan \left( \frac{\sqrt{\Delta}}{\delta_i}\right),
\eq
with
\bq
 \Delta = \mu_1 \mu_2 \mu_3 \mu_4,
 \;\;\;\;\;\;
 \delta_1 = -m_1^2 + m_2^2 + m_3^2,
 \;\;\;\;\;\;
 \delta_2 = m_1^2 - m_2^2 + m_3^2,
 \;\;\;\;\;\;
 \delta_3 = m_1^2 + m_2^2 - m_3^2.
\eq
Finally, the functions $b_1(m_1,m_2,m_3)$ and $b_0(m_1,m_2,m_3)$ are given by
\bq
 b_i\left(m_1,m_2,m_3\right) = 
          d_i\left(m_1,m_2,m_3\right) \ln \frac{m_1^2}{\mu^2} 
        + d_i\left(m_2,m_3,m_1\right) \ln \frac{m_2^2}{\mu^2} 
        + d_i\left(m_3,m_1,m_2\right) \ln \frac{m_3^2}{\mu^2},
\eq
with
\bq
 d_1\left(m_1,m_2,m_3\right) & = & 2 m_1^2 - m_2^2 - m_3^2 ,
 \nonumber \\
 d_0\left(m_1,m_2,m_3\right) & = & 2 m_1^4 - m_2^4 - m_3^4 -m_1^2 m_2^2 - m_1^2 m_3^2 + 2 m_2^2 m_3^2.
\eq

\section{Conclusions}
\label{sec:conclusions}

In this paper we discussed the analytic solution of the two-loop sunrise integral with unequal non-zero masses in two space-time dimensions.
We obtained the solution by solving an inhomogeneous second-order linear differential equation.
The homogeneous solutions involve complete elliptic integrals of the first kind.
The two-loop sunrise integral with non-zero masses is the simplest Feynman integral, which cannot be expressed in terms of
multiple polylogarithms.
The complications are related to an irreducible second-order differential operator.
We expect the methods discussed in this paper to be useful for other computations as well.

\subsection*{Acknowledgements}

Ch.B. thanks Dirk Kreimer's group at Humboldt University for support and hospitality.

\begin{appendix}

\section{Transformation to Legendre normal form}

In this appendix we discuss how to transform the general quartic elliptic curve
\bq
 y^2 & = & \left(x-x_1\right)\left(x-x_2\right)\left(x-x_3\right)\left(x-x_4\right)
\eq
to the cubic Legendre normal form
\bq
 w^2 & = & z \left(z-\lambda\right) \left(1-z\right).
\eq
Without loss of generality we can assume that
\bq
 x_1 < x_2 < x_3 < x_4.
\eq
The mapping
\bq
 z = \frac{\left(x_3-x_4\right)\left(x-x_1\right)}{\left(x_3-x_1\right)\left(x-x_4\right)},
 & &
 x = \frac{x_1\left(x_4-x_3\right)+x_4\left(x_3-x_1\right)z}{x_4-x_3+\left(x_3-x_1\right)z},
 \nonumber \\
 & &
 \frac{dx}{dz} = \frac{\left(x_3-x_1\right)\left(x_4-x_3\right)\left(x_4-x_1\right)}{\left[x_4-x_3+\left(x_3-x_1\right)z\right]^2}
\eq
sends the points $x_1$, $x_3$, $x_4$ to the points $0$, $1$ and infinity.
The point $x_2$ is mapped to
\bq
 \lambda & = &
 \frac{\left(x_2-x_1\right)\left(x_4-x_3\right)}{\left(x_3-x_1\right)\left(x_4-x_2\right)}.
\eq
We thus have
\bq
 \frac{dx}{y} & = &
 \frac{1}{\sqrt{\left(x_3-x_1\right)\left(x_4-x_2\right)}} \frac{dz}{w}
 =
 \frac{1}{\sqrt{\left(x_3-x_1\right)\left(x_4-x_2\right)}} \frac{dz}{\sqrt{z\left(1-z\right)\left(z-\lambda\right)}}
\eq
and with the notation of eq.~(\ref{def_complementary_modulus}) $k'=\sqrt{\lambda}$.
With this mapping we obtain for the integrals
\bq
 \int\limits_{x_2}^{x_3} \frac{dx}{y}
 & = &
 \frac{1}{\sqrt{\left(x_3-x_1\right)\left(x_4-x_2\right)}} \int\limits_\lambda^1 \frac{dz}{w}
 =
 \frac{2}{\sqrt{\left(x_3-x_1\right)\left(x_4-x_2\right)}} K\left(\sqrt{1-\lambda}\right),
 \nonumber \\
 \int\limits_{x_4}^{x_3} \frac{dx}{y}
 & = &
 \frac{1}{\sqrt{\left(x_3-x_1\right)\left(x_4-x_2\right)}} \int\limits_\infty^1 \frac{dz}{w}
 =
 \frac{2 i}{\sqrt{\left(x_3-x_1\right)\left(x_4-x_2\right)}} K\left(\lambda\right).
\eq

\section{Representation in terms of Lauricella functions}

The main focus of this paper is to represent the two-loop sunrise integral as an iterated integral 
in the spirit of section~\ref{sec:motivation}.
Besides iterated integrals a representation in terms of nested sums can also be useful.
In this appendix we briefly review a nested sum representation for the two-loop sunrise integral.
In $D$ space-time dimensions this representation is given in terms of Lauricella functions.
Unfortunately this representation does not allow a straightforward specialisation to two space-time dimensions.
This is due to spurious singularities which appear in intermediate results.
We will discuss this issue in the following.
We start with the two-loop sunrise integral in $D$ space-time dimensions.
\bq
 S\left( p^2 \right)
 & = &
 \left( \mu^2 \right)^{3-D}
 \int \frac{d^Dk_1}{i \pi^{\frac{D}{2}}} \frac{d^Dk_2}{i \pi^{\frac{D}{2}}}
 \frac{1}{\left(-k_1^2+m_1^2\right)\left(-k_2^2+m_2^2\right)\left(-\left(p-k_1-k_2\right)^2+m_3^2\right)}.
\eq
The first step is to use the Mellin-Barnes technique to write the propagators as
\bq
\label{mellin_barnes}
\frac{1}{\left(-k^2+m^2\right)}
 & = & 
 \frac{1}{2\pi i} \int d\sigma \; \Gamma\left(-\sigma\right)\Gamma\left(\sigma+1\right) \frac{\left(m^2\right)^\sigma}{\left(-k^2\right)^{\sigma+1}}.
\eq
The integration in eq.~(\ref{mellin_barnes}) 
is over a contour along the imaginary axis separating the poles of $\Gamma\left(-\sigma\right)$ and $\Gamma\left(\sigma+1\right)$.
The integration over the loop momenta $k_1$ and $k_2$ corresponds then to an integration for a massless sunrise integral and can be done easily.
Setting
\bq
 x_1 = \frac{m_1^2}{(-t)},
 \;\;\;\;\;\;
 x_2 = \frac{m_2^2}{(-t)},
 \;\;\;\;\;\;
 x_3 = \frac{m_3^2}{(-t)},
\eq
one arrives at the Mellin-Barnes representation
\bq
\lefteqn{
 S\left( t \right)
 = 
 \left(\frac{-t}{\mu^2}\right)^{D-3} \frac{1}{\left(2 \pi i \right)^3} \int d\sigma_1 \; d\sigma_2 \; d\sigma_3 \; 
 x_1^{\sigma_1} x_2^{\sigma_2} x_3^{\sigma_3}
 \Gamma\left(-\sigma_1\right) \Gamma\left(-\sigma_2\right) \Gamma\left(-\sigma_3\right)
} & &
 \nonumber \\
 & &
 \Gamma\left(-\sigma_1+\frac{D}{2}-1\right)
 \Gamma\left(-\sigma_2+\frac{D}{2}-1\right)\Gamma\left(-\sigma_3+\frac{D}{2}-1\right)
 \frac{\Gamma\left(\sigma_1+\sigma_2+\sigma_3+3-D\right)}{\Gamma\left(-\sigma_1-\sigma_2-\sigma_3+\frac{3}{2}D-3\right)}.
\nonumber
\eq
The next step is then to close the contours and to take residues.
For $|x_i|<1$ ($i=1,2,3$) we can close all contours to the right and we obtain a symmetric result in $x_1$, $x_2$ and $x_3$.
The result can be expressed in terms of the third Lauricella function in three variables, defined by
\bq
F_C(a_1,a_2;b_1,b_2,b_3;x_1,x_2,x_3)  
= \sum\limits_{m_1=0}^\infty \sum\limits_{m_2=0}^\infty \sum\limits_{m_3=0}^\infty 
\frac{(a_1)_{m_1+m_2+m_3} (a_2)_{m_1+m_2+m_3}}
{(b_1)_{m_1} (b_2)_{m_2} (b_3)_{m_3}}
\frac{x_1^{m_1}}{m_1!} \frac{x_2^{m_2}}{m_2!} \frac{x_3^{m_3}}{m_3!},
\eq
where $(a)_n= \Gamma(a+n)/\Gamma(a)$ denotes the Pochhammer symbol.
We obtain
\bq
\label{lauricella_result}
\lefteqn{
 S\left( t \right)
 = 
 \left(\frac{-t}{\mu^2}\right)^{D-3}
} & & \nonumber \\
 & &
 \times
 \left\{
       \frac{\Gamma\left(3-D\right) \Gamma\left(\frac{D}{2}-1\right)^3}{\Gamma\left(\frac{3}{2}D-3\right)}
       F_C(3-D, 4-\frac{3}{2}D; 2-\frac{D}{2}, 2-\frac{D}{2}, 2-\frac{D}{2};-x_1,-x_2,-x_3)  
 \right.
 \nonumber \\
 & &
 \left.
+
       \frac{\Gamma\left(2-\frac{D}{2}\right) \Gamma\left(1-\frac{D}{2}\right) \Gamma\left(\frac{D}{2}-1\right)^2}{\Gamma\left(D-2\right)}
 \left[
       F_C(3-D, 2-\frac{D}{2}; \frac{D}{2}, 2-\frac{D}{2}, 2-\frac{D}{2};-x_1,-x_2,-x_3) x_1^{\frac{D}{2}-1}
 \right. \right.
 \nonumber \\
 & &
 \left. \left.
+
       F_C(3-D, 2-\frac{D}{2}; 2-\frac{D}{2}, \frac{D}{2}, 2-\frac{D}{2};-x_1,-x_2,-x_3) x_2^{\frac{D}{2}-1}
 \right. \right.
 \nonumber \\
 & &
 \left. \left.
+
       F_C(3-D, 2-\frac{D}{2}; 2-\frac{D}{2}, 2-\frac{D}{2}, \frac{D}{2};-x_1,-x_2,-x_3) x_3^{\frac{D}{2}-1}
 \right]
 \right.
 \nonumber \\
 & & \left.
+
       \Gamma\left(1-\frac{D}{2}\right)^2
 \left[
       F_C(1, 2-\frac{D}{2}; \frac{D}{2}, \frac{D}{2}, 2-\frac{D}{2};-x_1,-x_2,-x_3) \left(x_1 x_2\right)^{\frac{D}{2}-1}
 \right. \right.
 \nonumber \\
 & &
 \left. \left.
+
       F_C(1, 2-\frac{D}{2}; \frac{D}{2}, 2-\frac{D}{2}, \frac{D}{2};-x_1,-x_2,-x_3) \left(x_1 x_3\right)^{\frac{D}{2}-1}
 \right. \right.
 \nonumber \\
 & &
 \left. \left.
+
       F_C(1, 2-\frac{D}{2}; 2-\frac{D}{2}, \frac{D}{2}, \frac{D}{2};-x_1,-x_2,-x_3) \left(x_2 x_3\right)^{\frac{D}{2}-1}
 \right]
 \right\}.
\eq
This expresses the sunrise integral in $D$ space-time dimensions as a linear combination of seven Lauricella functions.
This result can be found already in \cite{Berends:1993ee}, however eq.~(24) of \cite{Berends:1993ee} has a few typos.
For the convenience of the reader we give here the correct result.

Eq.~(\ref{lauricella_result}) expresses the sunrise integral in $D$ space-time dimensions as a linear combination of seven 
Lauricella functions.
We are interested in the result in two space-time dimensions. In two space-time dimensions, eq.~(\ref{lauricella_result})
is not yet particular useful, because the individual terms diverge in the $D\rightarrow 2$ limit due to the prefactors
$\Gamma(D/2-1)$ and $\Gamma(1-D/2)$.
If we set $D=2-2\eps$, then each of the seven terms has a Laurent series in $\eps$ with poles up to $1/\eps^2$.
The poles are spurious singularities and cancel in the sum.
We obtain a manifest finite result for two-space time dimensions by expanding in $\eps$.
In order to present the result, we introduce the Euler-Zagier sums \cite{Zagier}
\bq
 Z_1\left(n\right) = \sum\limits_{j=1}^n \frac{1}{j},
 & &
 Z_{11}\left(n\right) = \sum\limits_{j=1}^n \frac{1}{j} Z_1\left(j-1\right).
\eq
Then the result is given by
\bq
\label{expanded_lauricella_result}
\lefteqn{
 S\left( t \right)
 = 
 \frac{\mu^2}{(-t)}
 \sum\limits_{j_1=0}^\infty \sum\limits_{j_2=0}^\infty \sum\limits_{j_3=0}^\infty 
 \left( \frac{j_{123}!}{j_1! j_2! j_3!} \right)^2
 \left(-x_1\right)^{j_1} \left(-x_2\right)^{j_2} \left(-x_3\right)^{j_3}
} & & \nonumber \\
 & &
 \times
 \left\{ 
 12 Z_{11}\left(j_{123}\right) 
 + 6 Z_{1}\left(j_{123}\right) Z_{1}\left(j_{123}\right)
 - 8 Z_{1}\left(j_{123}\right) \left[ Z_{1}\left(j_{1}\right) + Z_{1}\left(j_{2}\right) + Z_{1}\left(j_{3}\right) \right]
 \right. \nonumber \\
 & & \left.
 +4 \left[ Z_{1}\left(j_{1}\right) Z_{1}\left(j_{2}\right) + Z_{1}\left(j_{2}\right) Z_{1}\left(j_{3}\right) + Z_{1}\left(j_{3}\right) Z_{1}\left(j_{1}\right) \right]
 + 2 \left[ 2 Z_{1}\left(j_{123}\right) - Z_{1}\left(j_{2}\right) - Z_{1}\left(j_{3}\right) \right] \ln x_1
 \right. \nonumber \\
 & & \left.
 + 2 \left[ 2 Z_{1}\left(j_{123}\right) - Z_{1}\left(j_{3}\right) - Z_{1}\left(j_{1}\right) \right] \ln x_2
 + 2 \left[ 2 Z_{1}\left(j_{123}\right) - Z_{1}\left(j_{1}\right) - Z_{1}\left(j_{2}\right) \right] \ln x_3
 \right. \nonumber \\
 & & \left.
 + \ln x_1 \ln x_2 + \ln x_2 \ln x_3 + \ln x_3 \ln x_1 
 \right\}.
\eq
Here we used the notation $j_{123}=j_1+j_2+j_3$.
This expresses the two-loop sunrise integral in two space-time dimensions as a five-fold sum.

\end{appendix}

\bibliography{/home/stefanw/notes/biblio}

\begin{thebibliography}{10}

\bibitem{Broadhurst:1993mw}
D.~J. Broadhurst, J.~Fleischer, and O.~Tarasov,
\newblock Z.Phys. {\bf C60}, 287 (1993), arXiv:hep-ph/9304303.

\bibitem{Berends:1993ee}
F.~A. Berends, M.~Buza, M.~B{\"o}hm, and R.~Scharf,
\newblock Z.Phys. {\bf C63}, 227 (1994).

\bibitem{Bauberger:1994nk}
S.~Bauberger, M.~B{\"o}hm, G.~Weiglein, F.~A. Berends, and M.~Buza,
\newblock Nucl.Phys.Proc.Suppl. {\bf 37B}, 95 (1994), arXiv:hep-ph/9406404.

\bibitem{Bauberger:1994by}
S.~Bauberger, F.~A. Berends, M.~B{\"o}hm, and M.~Buza,
\newblock Nucl.Phys. {\bf B434}, 383 (1995), arXiv:hep-ph/9409388.

\bibitem{Caffo:1998du}
M.~Caffo, H.~Czyz, S.~Laporta, and E.~Remiddi,
\newblock Nuovo Cim. {\bf A111}, 365 (1998), arXiv:hep-th/9805118.

\bibitem{Laporta:2004rb}
S.~Laporta and E.~Remiddi,
\newblock Nucl. Phys. {\bf B704}, 349 (2005), hep-ph/0406160.

\bibitem{Groote:2005ay}
S.~Groote, J.~G. K{\"o}rner, and A.~A. Pivovarov,
\newblock Annals Phys. {\bf 322}, 2374 (2007), arXiv:hep-ph/0506286.

\bibitem{Groote:2012pa}
S.~Groote, J.~K{\"o}rner, and A.~Pivovarov,
\newblock Eur.Phys.J. {\bf C72}, 2085 (2012), arXiv:1204.0694.

\bibitem{Bailey:2008ib}
D.~H. Bailey, J.~M. Borwein, D.~Broadhurst, and M.~L. Glasser,
\newblock (2008), arXiv:0801.0891.

\bibitem{MullerStach:2011ru}
S.~M{\"u}ller-Stach, S.~Weinzierl, and R.~Zayadeh,
\newblock Commun. Num. Theor. Phys. {\bf 6}, 203 (2012), arXiv:1112.4360.

\bibitem{Berends:1997vk}
F.~A. Berends, A.~I. Davydychev, and N.~Ussyukina,
\newblock Phys.Lett. {\bf B426}, 95 (1998), arXiv:hep-ph/9712209.

\bibitem{Davydychev:1999ic}
A.~I. Davydychev and V.~A. Smirnov,
\newblock Nucl. Phys. {\bf B554}, 391 (1999), arXiv:hep-ph/9903328.

\bibitem{Caffo:1999nk}
M.~Caffo, H.~Czyz, and E.~Remiddi,
\newblock Nucl. Phys. {\bf B581}, 274 (2000), arXiv:hep-ph/9912501.

\bibitem{Caffo:2001de}
M.~Caffo, H.~Czyz, and E.~Remiddi,
\newblock Nucl. Phys. {\bf B611}, 503 (2001), arXiv:hep-ph/0103014.

\bibitem{Groote:2000kz}
S.~Groote and A.~Pivovarov,
\newblock Nucl.Phys. {\bf B580}, 459 (2000), arXiv:hep-ph/0003115.

\bibitem{Onishchenko:2002ri}
A.~Onishchenko and O.~Veretin,
\newblock Phys. Atom. Nucl. {\bf 68}, 1405 (2005), arXiv:hep-ph/0207091.

\bibitem{Bauberger:1994hx}
S.~Bauberger and M.~Bohm,
\newblock Nucl.Phys. {\bf B445}, 25 (1995), arXiv:hep-ph/9501201.

\bibitem{Caffo:2002ch}
M.~Caffo, H.~Czyz, and E.~Remiddi,
\newblock Nucl. Phys. {\bf B634}, 309 (2002), arXiv:hep-ph/0203256.

\bibitem{Pozzorini:2005ff}
S.~Pozzorini and E.~Remiddi,
\newblock Comput. Phys. Commun. {\bf 175}, 381 (2006), arXiv:hep-ph/0505041.

\bibitem{Caffo:2008aw}
M.~Caffo, H.~Czyz, M.~Gunia, and E.~Remiddi,
\newblock Comput. Phys. Commun. {\bf 180}, 427 (2009), arXiv:0807.1959.

\bibitem{Paulos:2012nu}
M.~F. Paulos, M.~Spradlin, and A.~Volovich,
\newblock JHEP {\bf 1208}, 072 (2012), arXiv:1203.6362.

\bibitem{CaronHuot:2012ab}
S.~Caron-Huot and K.~J. Larsen,
\newblock JHEP {\bf 1210}, 026 (2012), arXiv:1205.0801.

\bibitem{Tarasov:1996br}
O.~V. Tarasov,
\newblock Phys. Rev. {\bf D54}, 6479 (1996), hep-th/9606018.

\bibitem{Tarasov:1997kx}
O.~V. Tarasov,
\newblock Nucl. Phys. {\bf B502}, 455 (1997), hep-ph/9703319.

\bibitem{MullerStach:2012mp}
S.~M{\"u}ller-Stach, S.~Weinzierl, and R.~Zayadeh,
\newblock (2012), arXiv:1212.4389.

\bibitem{Laporte:2012hv}
G.~Laporte and J.~Walcher,
\newblock SIGMA {\bf 8}, 056 (2012), arXiv:1206.1787.

\bibitem{fettis:1970:rmr}
H.~E. Fettis,
\newblock SIAM J. Math. Anal. {\bf 1}, 524 (1970).

\bibitem{Davydychev:1995mq}
A.~I. Davydychev and J.~Tausk,
\newblock Phys.Rev. {\bf D53}, 7381 (1996), arXiv:hep-ph/9504431.

\bibitem{Ussyukina:1993jd}
N.~I. Ussyukina and A.~I. Davydychev,
\newblock Phys. Lett. {\bf B298}, 363 (1993).

\bibitem{Lu:1992ny}
H.~J. Lu and C.~A. Perez,
\newblock SLAC-PUB-5809.

\bibitem{Bern:1997ka}
Z.~Bern, L.~Dixon, D.~A. Kosower, and S.~Weinzierl,
\newblock Nucl. Phys. {\bf B489}, 3 (1997), hep-ph/9610370.

\bibitem{Zagier}
D.~Zagier,
\newblock First European Congress of Mathematics, Vol. II, Birkhauser, Boston ,
  497 (1994).

\end{thebibliography}
\bibliographystyle{/home/stefanw/latex-style/h-physrev5}

\end{document}